\documentclass[12pt,nohyper]{JHEP}          
\usepackage{epsfig}
\usepackage{amssymb}
\renewcommand{\d}     {\mbox{d}}

\newcommand  {\vev}[1]{\left\langle #1 \right\rangle}

\newcommand{\id}{{1\!\!1}} 
\newcommand\jp   [3]{{{\it J.\ Phys.\ }{\bf #1} (#2) #3}}
\newcommand\pl  [3]{{{\it Phys.\ Lett.\ }{\bf #1} (#2) #3}}
\font\mybbs=msbm10 at 9pt
\def\bbs#1{\hbox{\mybbs#1}}
\font\mybb=msbm10 at 12pt
\def\bb#1{\hbox{\mybb#1}}

\def\IC{{\bb C}}

\def\IZs{{\bbs Z}}

\def\rref#1{(\ref{#1})}

\newcommand {\bd}{\begin{displaymath}}
\newcommand {\ed}{\end{displaymath}}
\newcommand {\eq}{\begin{equation}}
\newcommand {\beq}{\begin{equation}}
\newcommand {\eeq}{\end{equation}}
\newcommand {\beqa}{\begin{eqnarray}}
\newcommand {\eeqa}{\end{eqnarray}}
\newcommand {\n}{\nonumber \\}

\newcommand {\eex}{\mbox{e}}

\newcommand {\ddd}{\mbox{d}}

\newcommand {\del}{\partial}

\newcommand {\defeq}{\stackrel{\rm def}{=}}

\newcommand{\tr}{\mathop{\mathrm{tr}}\nolimits}

\def\bea{\begin{eqnarray}}
\def\eea{\end{eqnarray}}
\def\be{\begin{equation}}
\def\ee{\end{equation}}

\def\d{\partial}

\newcommand{\rf}[1]{(\ref{#1})}

\title{Noncommutative String Worldsheets from Matrix Models}
\author{K.N.\ Anagnostopoulos
\\Department of Physics, University of Crete,\\
P.O.Box 2208, GR-710 03 Heraklion, Crete, Greece\\
Email: {\tt konstant@physics.uoc.gr}}
\author{J.\ Nishimura\footnote{On leave of absence from Department
of Physics, Nagoya University, Furo-cho, Chikusa-ku, Nagoya 464-8602, 
Japan.}\\
The Niels Bohr Institute, \\
Blegdamsvej 17, DK-2100 Copenhagen \O , Denmark\\
E-mail: {\tt nisimura@nbi.dk} }
\author{P.\ Olesen\\
The Niels Bohr Institute, \\
Blegdamsvej 17, DK-2100 Copenhagen \O , Denmark\\
E-mail: {\tt polesen@nbi.dk} }

\abstract{ 
We study dynamical effects of introducing noncommutativity on string
worldsheets by using a matrix model obtained from the zero-volume
limit of four-dimensional SU($N$) Yang-Mills theory.  Although the
dimensionless noncommutativity parameter is of order $1/N$, its effect
is found to be non-negligible even in the large $N$ limit due to the
existence of higher Fourier modes.  We find that the Poisson bracket
grows much faster than the Moyal bracket as we increase $N$, which
means in particular that the two quantities do not coincide in the
large $N$ limit. The well-known instability of bosonic worldsheets 
due to long spikes is shown to be cured by the noncommutativity.
The extrinsic geometry of the worldsheet is described by
a crumpled surface with a large Hausdorff dimension.
} 

\preprint{NBI-HE-00-43}

\begin{document}
\section{Introduction}\label{intro}
\setcounter{footnote}{0}
It is well known that the traditional bosonic worldsheet theories
described for example by the Nambu-Goto action, the Polyakov action
and the Schild action are not well-defined 
non-perturbatively, as was first observed in the large dimension limit
by Alvarez \cite{alvarez}. 
This conclusion has been confirmed more directly by
the dynamical triangulation approach \cite{dt,poles},
where the discretized worldsheet embedded in the target space
degenerates to long spikes, and hence one cannot view it
as a proper approximation of a continuous worldsheet.
It is therefore of interest to inquire whether it is
possible to have different types of string models, where the worldsheet 
is well-defined in the sense that it does not degenerate into long
spikes. Some attempts have been made to make the string more ``stiff'' by
the introduction of extrinsic curvature \cite{ec}, but it is not clear whether
this will work non-perturbatively \cite{exx}.

In the present paper we have investigated whether the introduction
of noncommutativity on the worldsheet changes the situation.
We study such a system by using a matrix model with $N\times N$ hermitian
matrices $A_\mu$ with $\mu=1,2,3,4.$ 
Corresponding to each matrix one can construct a 
field $X_\mu (\sigma)$ by the Weyl transformation. Here
$\sigma$ stands for the two discrete variables $\sigma_1$ and $\sigma_2$,
representing the worldsheet coordinates. 
Then there exists the following relation between the partition functions
in the matrix model and in the corresponding Weyl-transformed model,
\beq
\int \ddd A_\mu
  \exp\left(\frac{1}{4 g^2}{\rm
  Tr}[A_\mu,A_\nu]^2\right) = 
\int {\cal D} X_\mu(\sigma) \, \eex ^{-S} \ ,
\label{1}
\eeq
where the action $S$ for the latter is given by
\be
S= - \frac{1}{4 g^2 N}
\sum_\sigma
\Bigl( X_\mu (\sigma) \star
X_\nu (\sigma)
-X_\nu (\sigma)\star X_\mu (\sigma) \Bigr )^2 \ .
\label{nonc}
\ee
Here $\star$ denotes the star product
with the dimensionless
noncommutative parameter being of order $1/N$.
The partition function on the left hand side of eq.~(\ref{1}) 
was shown to be finite for $N=3,4,5$ numerically by Ref.~\cite{KS}.
Monte Carlo simulations up to $N=256$ \cite{HNT} further suggests that
the statement extends
\footnote{This has been proved rigorously 
in a preprint \cite{Austing:2001bd},
which appeared while our paper was being revised.}
to $N = \infty$.
This means that the corresponding Weyl-transformed model is also 
well-defined in the large $N$ limit.

The partition function on the right hand side of eq.~(\ref{1}) 
defines a
noncommutative two-dimensional field theory, which is invariant under the
star-unitary transformation,
\be
X_\mu (\sigma)\rightarrow g^*(\sigma)\star X_\mu (\sigma)\star 
g(\sigma) \ ,~~
{\rm with}~~ g^*(\sigma )\star g(\sigma)=g(\sigma )\star g^*(\sigma)=1 \ .
\ee
The action (\ref{nonc}) resembles the Schild action
\be
S_{\rm Schild} = \frac{1}{g^2 N}
\int \ddd ^2 \sigma \left(\frac{\d X_\mu}{\d\sigma_1}\frac{\d X_\nu}
{\d\sigma_2}-\frac{\d X_\nu}{\d\sigma_1}\frac{\d X_\mu}
{\d\sigma_2}\right)^2 \ ,
\label{sch}
\ee
the difference being that the Poisson bracket has
been replaced by the Moyal bracket in the noncommutative case.
It has been pointed out long time ago \cite{longtime} that if the higher
modes in the mode expansion of $X_\mu (\sigma)$ can be neglected, the 
star-Schild action in (\ref{nonc}) 
reduces to the Schild action (\ref{sch}) in the large $N$ limit.
The issue is also relevant when one considers matrix models
as a nonperturbative definition
of M-theory and type IIB superstring theory \cite{BFSS,IKKT}.
Of course, whether the higher modes can really be neglected or not
is a non-trivial dynamical question.
One of the main results of
this paper is that for the bosonic model we consider,
the Poisson and the Moyal brackets are very
different due to the effect of the higher modes,
and that this effect increases with increasing $N$.
To our knowledge, this is the first time that the agreement
(or disagreement) of the two actions in the large $N$ limit
is discussed in a dynamical context.

Recently the Schild action has been investigated non-perturbatively from the 
point of view of dynamical triangulation \cite{poles}. The result is
similar to the one obtained for the Nambu-Goto action, namely that it is
dynamically favorable to have a ``surface'' which degenerates to
long spikes, and hence the
notion of a worldsheet looses its meaning\footnote{A similar result is not 
valid in the supersymmetric case, where the worldsheet exists if the fermions
are coupled to a bosonic Schild action \cite{poles}. However, in the present 
paper we shall discuss the bosonic case only.}. 

We shall take a different approach and ask whether 
the star-Schild action in (\ref{nonc}) 
provides a bosonic string theory
with a well-defined worldsheet. 
The effects of noncommutativity 
are found to be drastic.
The average link length is finite and it is 
observed to be considerably smaller than 
the average extent of the surface,
which is in sharp contrast \cite{alvarez,poles} to the conclusion for 
the Schild action (\ref{sch}) that
the worldsheet becomes completely unstable due to long spikes.
This is not a contradiction since
the Poisson and Moyal brackets are indeed found to be quite different
for the theory under consideration.
On the other hand, the extrinsic geometry 
of the worldsheet for the star-Schild action
is described by a crumpled surface 
with a large Hausdorff dimension.

The matrix model on the left hand side of eq.~(\ref{1}) 
can be regarded as the hermitian matrix version
of the Eguchi-Kawai model \cite{EK}.
(See Refs.~\cite{GrossKitazawa} and \cite{contEK} 
for the hermitian matrix version of 
the Eguchi-Kawai model with 
quenching \cite{quench} or twists \cite{GO}, respectively.)
Large $N$ scaling of SU($N$) invariant correlation functions
was obtained analytically to all orders of the $1/D$ expansion \cite{HNT},
and the results were reproduced by Monte Carlo simulations \cite{HNT,AABHN}.
In particular, the one-point function of a Wilson loop was observed to
obey the area law, which suggested that the model is actually equivalent to
large $N$ gauge theory for a finite region of scale
even {\em without} quenching \cite{quench} or twists \cite{GO}.
The noncommutative worldsheet theory studied in the present paper
is therefore related to the large $N$ QCD string 
(For a recent review, see \cite{tHooft}.). 

The organization of this paper is as follows.
In Section \ref{review} 
we briefly review the map from matrices to noncommutative fields.
In Section \ref{matrixworldsheet} we interpret the matrix model as 
noncommutative worldsheet theory. We then discuss
the star-unitary invariance and the 
important question of gauge fixing. 
In Section \ref{results} the results are presented. 
Section \ref{conclusions} contains the summary and discussions.
In an Appendix we make some comments on the relationship
between the star-unitary invariance and 
the area-preserving diffeomorphisms.


\section{From matrices to noncommutative fields}
\label{review}
\setcounter{equation}{0}

In order to derive the equivalence (\ref{1}) between the matrix model
and the noncommutative worldsheet theory,
we briefly review the one-to-one correspondence
between matrices and noncommutative fields.
Most of the results in this
Section are known in the literature 
(see e.g., \cite{BM,AMNS}),
but are given in order to fix the notation.

Throughout this paper, we assume that $N$ is odd\footnote{We
expect that the large $N$ dynamics of the matrix model 
on the l.h.s.\ of (\ref{1})
is independent of whether $N$ is even or odd.
This has been also checked numerically for various SU($N$) 
invariant quantities.
However, the one-to-one correspondence
between the matrices and noncommutative fields works rigorously
only for odd $N$. 
}.
We first introduce the 't Hooft-Weyl algebra
\beq
\Gamma_1 \Gamma_2 = \omega \,\Gamma_2 \Gamma_1 \ ,
\eeq
where $\omega = \eex^{4\pi i/N}$.
It is known that the representation of
$\Gamma_i$ using $N \times N$ unitary matrices 
is unique up to unitary 
transformation $\Gamma_i \rightarrow U \Gamma_i U^\dag$,
where $U \in \mbox{SU}(N)$. 
An explicit representation can be given by
\beq
\Gamma _1=
\pmatrix{0&1& & &0\cr &0&1& & \cr& &\ddots&\ddots& \cr& & &\ddots&1\cr 1& &
& &0\cr}~~~~~~,~~~~~~
\Gamma_2=\pmatrix{1& & & & \cr &\omega& & & \cr&
&\omega ^ 2 & & \cr& & &\ddots& \cr & & & & \omega ^{(N-1)}\cr} \ .
\label{clockshift}\eeq
Then we construct $N \times N$ unitary matrices
\beq
J_{k} \defeq
(\Gamma_1)^{k_1} (\Gamma_2)^{k_2}
{}~\eex ^{- 2 \pi i  k_1 k_2 /N} \ ,
\label{defJ}
\eeq
where $k$ is a 2d integer vector and 
the phase factor $\eex ^{- 2 \pi i  k_1 k_2 /N}$ 
is included so that
\beq
J_{-k}=(J_{k}) ^ \dag \ .
\label{Jconjg}
\eeq
Since $(\Gamma_i)^N=1$, the matrix 
$J_{k}$ is periodic with respect to
$k_i$ with period $N$, 
\beq
J_{k_1+ N,k_2} = J_{k_1,k_2+N} = J_{k_1,k_2} \ .
\eeq
We then define the $N \times N$ matrices
\beq
\Delta (\sigma) \defeq  \sum _{k_i \in \IZs _N} 
J_{k}~\eex ^{ 2 \pi i  k  \cdot \sigma / N } \ ,
\label{defDelta}
\eeq
labelled by a 2d integer vector $\sigma$.
Note that $\Delta (\sigma)$ is hermitian 
due to the property (\ref{Jconjg}).
It is periodic with respect to $\sigma_i$ with period
$N$,
\beq
\Delta (\sigma_1 + N , \sigma_2)
= \Delta (\sigma_1 , \sigma_2 + N)
= \Delta (\sigma_1 , \sigma_2) \ .
\label{periodicity}
\eeq
It is easy to check that $\Delta (\sigma)$ 
possesses the following properties.
\beqa
\label{trDelta}
\tr \Delta (\sigma) &=& N  \\
\label{sumDelta}
\sum _{\sigma_i \in \IZs _N} \Delta (\sigma) &=& N^2 {\bf 1}_N \\
\frac{1}{N} \tr \Bigl( \Delta  (\sigma)  \Delta (\sigma') \Bigr) &=& 
N^2 \delta_{\sigma,\sigma ' (\mbox{\scriptsize mod~}N)} \ .
\label{orthogonal}
\eeqa

Now let us consider
gl($N$,$\IC$),
the linear space of $N \times N$ 
complex matrices. 
The inner product can be defined by 
$\tr (M_1 ^\dag M_2)$,
where
$M_1$,$M_2$$\in$
gl($N$,$\IC$).
Eq.~(\ref{orthogonal}) implies that 
$\Delta (\sigma)$ are mutually orthogonal and hence linearly independent.
Since the dimension of the linear space gl($N$,$\IC$) is $N^2$ and
there are $N^2$ $\Delta (\sigma)$'s that are linearly independent,
$\Delta (\sigma)$ actually forms an orthogonal basis of gl($N,\IC$).
Namely, any $N \times N$ complex matrix $M$
can be written as
\beq
M  = \frac{1}{N^2}\sum _{\sigma_i \in \IZs _N} f(\sigma) 
\Delta (\sigma) \ ,
\label{complete}
\eeq
where $f (\sigma)$ is a complex-valued function on the 2d discretized
torus obeying periodic boundary conditions.
Using orthogonality (\ref{orthogonal}),
one can invert (\ref{complete}) as
\beq
f(\sigma)= \frac1N\,\tr\Bigl(M \,\Delta(\sigma)\Bigr) \ .
\label{invert}
\eeq
The fact that (\ref{complete}) and (\ref{invert}) hold
for an arbitrary $N \times N$ complex matrix $M$ implies that
$\Delta (\sigma)$ satisfies
\beq
\frac{1}{N^2}
\sum _{\sigma_i \in \IZs _N} \Delta _{ij} (\sigma)  \Delta_{kl} (\sigma) 
= N \delta_{il} \delta _{jk} \ .
\eeq
Note also that using (\ref{trDelta}) with (\ref{complete}) or
using (\ref{sumDelta}) with (\ref{invert}),
one obtains
\beq
\frac{1}{N}\tr M  = 
\frac{1}{N^2}\sum _{\sigma_i \in \IZs _N} f(\sigma)  \ .
\label{trace}
\eeq


We define the ``star-product'' of two functions 
$f _1(\sigma)$ and $f _2(\sigma)$ by
\beq
f_1(\sigma) \star f_2(\sigma)
\defeq
\frac{1}{N} \tr (M_1 M_2 \Delta (\sigma) ) \ ,
\label{stardef}
\eeq
where 
\beq
M _\alpha  =  \frac{1}{N^2}\sum _{\sigma_i \in \IZs _N} f_\alpha (\sigma) 
\Delta (\sigma)\mbox{~~~~~~~}\alpha=1,2 \ .
\label{ftoM}
\eeq
The star-product can be 
written explicitly in terms of $f_\alpha (\sigma)$ as
\beq
f_1(\sigma) \star f_2(\sigma)
= \frac{1}{N^2} \sum _{\xi _i \in \IZs_N}
\sum _{\eta _i \in \IZs_N}
 f_1(\xi) f_2(\eta)
\eex ^{-2 \pi i \epsilon _{jk} 
(\xi_j - \sigma _j  ) (\eta_k - \sigma _k )/N} \ ,
\label{starexplicit}
\eeq
where $\epsilon _{ij}$ is an antisymmetric tensor with 
$\epsilon _{12}  = 1$.
This formula can be derived in the following way.
Substituting (\ref{ftoM}) into (\ref{stardef})
and using the definition (\ref{defDelta}) of 
$\Delta(\sigma)$, one obtains
\beq
f_1(\sigma) \star f_2(\sigma)
= \left( \frac{1}{N^2} \right) ^2 
\sum _{\xi \eta}  \sum_{kpq} 
 f_1(\xi) f_2(\eta)
\eex ^{2 \pi i (p\cdot \xi + q\cdot \eta + k \cdot \sigma)}
 \frac{1}{N} \tr (J_{p}J_{q}J_{k}) \ .
\label{intermed}
\eeq
{}From the definition of $J_{k}$, one easily finds that
\beq
 \frac{1}{N} \tr (J_{p}J_{q}J_{k})
= \eex ^{ 2\pi i \epsilon _{ij} p_i q_j /N } 
\delta_{k+p+q,0(\mbox{\scriptsize mod~}N)} \ .
\eeq
Integration over $k$, $p$ and $q$
in (\ref{intermed}) yields eq.~(\ref{starexplicit}).

In order to confirm that
the star-product (\ref{starexplicit}) is a proper
discretized version of the usual star-product in the continuum,
we rewrite it in terms of Fourier modes.
We make a Fourier mode expansion of
$f_\alpha (\sigma)$ as
\beqa
f_\alpha (\sigma) &=& \sum _{k _i \in \IZs _N} 
\tilde{f}_\alpha(k) ~
\eex ^{2 \pi i k\cdot \sigma/N} \\
\tilde{f}_\alpha (k) &=& 
\frac{1}{N^2} \sum _{\sigma_i \in \IZs _N} f_\alpha(\sigma) ~
\eex ^{- 2 \pi i k\cdot \sigma/N} \ .
\eeqa
Integrating (\ref{intermed}) over $k$, $\xi$ and $\eta$,
one obtains
\beq
f_1(\sigma) \star f_2(\sigma)
= \sum _{p _i \in \IZs_N}
\sum _{q _i \in \IZs_N}
 \tilde{f}_1(p) \tilde{f}_2(q) ~
\eex ^{2 \pi i \epsilon _{jk} p _j q _k/N} 
\eex ^{2 \pi i (p+q) \cdot \sigma /N} \ ,
\label{starFourier}
\eeq
which can be compared to the usual
definition of the star-product in the continuum
\beq
f_1(\sigma) \star f_2(\sigma) = f_1(\sigma) \exp 
\Bigl( i \frac{1}{2} \theta \epsilon_{ij}
\overleftarrow{\del_i}
\overrightarrow{\del_j}
\Bigr) f_2(\sigma)  \ .
\eeq 
In the present case, 
the noncommutativity parameter $\theta$ is of order $1/N$,
and therefore the star-product reduces to the ordinary product
in the large $N$ limit
if $f_\alpha(\sigma)$ contains 
only lower Fourier modes ($p _j, q_j  \ll \sqrt{N}$).


It is obvious from the definition (\ref{stardef})
that the algebraic properties of the star-product are exactly 
those of the matrix product.
Namely, it is associative but not commutative.
Note also that due to (\ref{trace}), summing
a function $f(\sigma)$ over $\sigma$
corresponds to taking the trace of the corresponding matrix $M$.
Therefore, 
\beq
\sum _{\sigma_i \in \IZs _N} 
f_1(\sigma) \star f_2(\sigma) \star \cdots \star f_n(\sigma)
\eeq
is invariant under cyclic permutations of $f_\alpha (\sigma)$.
What is not obvious solely from the algebraic properties is that
\beq
\sum _{\sigma_i \in \IZs _N} 
f_1(\sigma) \star f_2(\sigma) 
= \sum _{\sigma_i \in \IZs _N} 
f_1(\sigma) f_2(\sigma)  \ ,
\eeq
which can be shown by using the definition
(\ref{stardef}) with eq.~(\ref{sumDelta}).


For later convenience, let us define the Moyal bracket by
\beqa
\{ \!\! \{ f_1(\sigma) ,  f_2(\sigma) \} \!\! \} 
&\defeq& i \, \frac{N}{4 \pi} 
\Bigl( f_1(\sigma) \star f_2(\sigma) -
f_2(\sigma) \star f_1(\sigma) \Bigr) \n
&=& - \frac{N}{2 \pi}   \sum _{pq}
 \tilde{f}_1(p) \tilde{f}_2(q)
\sin \left( \frac{2 \pi  \epsilon _{jk} p _j q _k}{N} \right) 
\eex ^{2 \pi i (p+q) \cdot \sigma /N}  \ .
\label{Moyal}
\eeqa
We also define the Poisson bracket
on a discretized worldsheet.
Namely when we define the Poisson bracket
\beq
\{ f_1 (\sigma) ,f_2 (\sigma) \} \defeq
\epsilon _{ij} 
\frac{\del f_1 (\sigma)}{\del \sigma _i}
\frac{\del f_2 (\sigma)}{\del \sigma _j} \ ,
\label{latticePoisson}
\eeq
we assume that the derivatives are given by the lattice derivatives
\beq
\frac{\del f (\sigma)}{\del \sigma _i}
= 
\frac{1}{2 a} \Bigl( f(\sigma + \hat{i} ) - f(\sigma - \hat{i} ) \Bigr) \ ,
\eeq
where $a = 2\pi /N$ is the lattice spacing.
The Poisson bracket thus defined can be written in terms of
Fourier modes as 
\beq
\{ f_1 (\sigma) ,f_2 (\sigma) \} 
= - \left( \frac{N}{2 \pi} \right)^2  
\sum _{pq}
 \tilde{f}_1(p) \tilde{f}_2(q)
\epsilon _{jk}
\sin \left( \frac{2 \pi   p _j }{N} \right) 
\sin \left( \frac{2 \pi   q _k}{N} \right) 
\eex ^{2 \pi i (p+q) \cdot \sigma /N}  \ .
\label{PoissonK}
\eeq
Note that the appearance of sines is due to
discretization of the worldsheet.
The Moyal bracket (\ref{Moyal}) and the Poisson bracket
(\ref{PoissonK}) agree in the large $N$ limit
if nonvanishing Fourier modes are those with $p_j,q_j \ll \sqrt{N}$.

\section{Matrix model as noncommutative worldsheet theory}
\label{matrixworldsheet}

Let us proceed to the derivation of the equivalence (\ref{1})
between the matrix model and the noncommutative worldsheet theory.
We start from the matrix model with the action
\beq
S = - \frac{1}{4 g^2} \tr ( [A_\mu, A_\nu] ^2 )  \ ,
\label{action}
\eeq
which can be obtained from the zero-volume limit of 
$D$-dimensional SU($N$) Yang-Mills theory\footnote{For $D=10$,
the action (\ref{action}) is just the bosonic part of the 
IIB matrix model \cite{IKKT}.
The dynamical aspects of this kind of matrix models for various $D$ 
with or without supersymmetry have been studied 
by many authors \cite{KS,HNT,AABHN,matrixmodels} both numerically 
and analytically.}.
The indices $\mu$ run from 1 to $D$.

As we have done in (\ref{complete}),
we write the $N \times N$ hermitian matrices $A_\mu$
in (\ref{action}) as
\beq
A _\mu  = \frac{1}{N^2}\sum _{\sigma_i \in \IZs _N} X_\mu (\sigma) 
\Delta (\sigma) \ ,
\label{mavsfu}
\eeq
where $X_\mu (\sigma)$ is a field on the discretized 2d torus
obeying periodic boundary conditions.
Since $A_\mu$ is hermitian, $X_\mu(\sigma)$ is real, due to 
the hermiticity of $\Delta (\sigma)$.
Eq.~\rref{mavsfu} can be inverted as
\beq
X_\mu(\sigma)=\frac1N\,\tr\Bigl(A_\mu\,\Delta(\sigma)\Bigr) \ .
\label{Uxinv}
\eeq
We regard $\sigma$ as (discretized) 
worldsheet coordinates and $X_\mu(\sigma)$ as the embedding 
function of the worldsheet into the target space.

Using the map discussed in the previous section we 
can rewrite the action (\ref{action})
in terms of $X_\mu (\sigma)$ as\footnote{While this work was being
completed, we received a preprint \cite{Kitsune} where
the worldsheet theory (\ref{Moyalaction}) is discussed
from a different point of view.}
\beqa
S &=& 
 - \frac{1}{4 g^2 N} \sum_{\sigma_i \in \IZs _N}
\Bigl( X_\mu(\sigma) \star X_\nu(\sigma) -
X_\nu(\sigma) \star X_\mu(\sigma) \Bigr) ^2\nonumber  \\
&=& 
 \frac{1}{g^2 N} \left( \frac{2 \pi}{ N} \right) ^2
 \sum_{\sigma_i \in \IZs _N}
 \{ \!\! \{  X_\mu(\sigma), X_\nu(\sigma) \} \!\! \}  ^2  \ .
\label{Moyalaction}
\eeqa
If $X_\mu (\sigma)$ are sufficiently smooth functions of $\sigma$,
the Moyal bracket can be replaced by 
the Poisson bracket, and therefore we obtain
the Schild action
\beq
 \label{Schild}
 S_{\rm Schild} = 
 \frac{1}{g^2 N} \left( \frac{2 \pi}{ N} \right) ^2
 \sum_{\sigma_i \in \IZs _N}
 \{   X_\mu(\sigma), X_\nu(\sigma) \}   ^2  \ .
\label{Schildaction}
\eeq

Let us discuss the symmetry of the theory
(\ref{Moyalaction}), which shall be important in our analysis.
For that we recall that the model (\ref{action}) is invariant under
the SU($N$) transformation
\beq
\label{gauge}
A_\mu \rightarrow g \, A_\mu \, g^\dag \ .
\eeq
{}From the matrix-field correspondence described in the previous section,
one easily finds that the field $X_\mu (\sigma)$ 
defined through (\ref{Uxinv}) transforms as
\beq
X_\mu (\sigma) \rightarrow 
g(\sigma) \star X_\mu (\sigma) \star g(\sigma)^*  \ ,
\label{stargaugetr}
\eeq
where $g(\sigma)$ is defined by
\beq
g(\sigma)=\frac1N\,\tr\Bigl(g\,\Delta(\sigma)\Bigr) \ .
\eeq
The fact that $g \in \mbox{SU}(N)$ implies
that $g(\sigma)$ is star-unitary;
\beq
g(\sigma)^* \star g(\sigma)
= g(\sigma) \star g(\sigma)^* = 1 \ .
\eeq
The action (\ref{Moyalaction}) is invariant under
the star-unitary transformation (\ref{stargaugetr}) as it should.
We shall refer to this invariance as 
`gauge' degrees of freedom in what follows.

Even if $X_\mu (\sigma)$
is a smooth function of $\sigma$ for a particular choice of gauge,
it can be made rough by making a rough star-unitary transformation.
Let us quote an analogous situation in lattice gauge theory.
In the weak coupling limit, 
the configurations can be made
very smooth by a proper choice of the gauge.
However, without gauge fixing, they are as rough as could be
due to the unconstrained gauge degrees of freedom.
Similarly when we discuss the smoothness of $X_\mu (\sigma)$,
we should subtract the roughness due to the 
gauge degrees of freedom appropriately.
Therefore, a natural question one should ask is 
whether there exists at all a gauge choice that
makes $X_\mu (\sigma)$ relatively smooth functions of $\sigma$.

In order to address this question,
we specify a gauge-fixing condition
by first defining the roughness of the worldsheet configuration
$X_\mu (\sigma)$ and
then choosing a gauge so that the roughness is minimized.
A natural definition of roughness is
\beq
I  = \frac{1}{2 N^2} \sum _{\sigma_i \in \IZs _N} \sum_{j=1}^{2} 
\Bigl( X_\mu (\sigma + \hat{j}) - X_\mu (\sigma) \Bigr)^2  \ ,
\label{defI}
\eeq
which is Lorentz invariant.
The gauge fixing is analogous to the Landau gauge in gauge theories.
The roughness functional (\ref{defI})
can be written conveniently 
in terms of $A_\mu$ as (See Appendix 
\ref{append_rough} for the derivation.)
\beq
\label{roughness}
I  = \frac{1}{2 N}  \sum _{IJ} 
\left[ 4 \sin ^2 \frac{ \pi (I-J)}{N} |(A_\mu)_{IJ}|^2
+ \left| (A_\mu)_{IJ} 
- (A_\mu)_{I+\frac{N-1}{2}, J+\frac{N-1}{2}}  \right|^2
\right]   \ .
\label{conv}
\eeq

\section{Results}
\label{results}

\FIGURE[hbt]{
    \includegraphics[height=8cm]{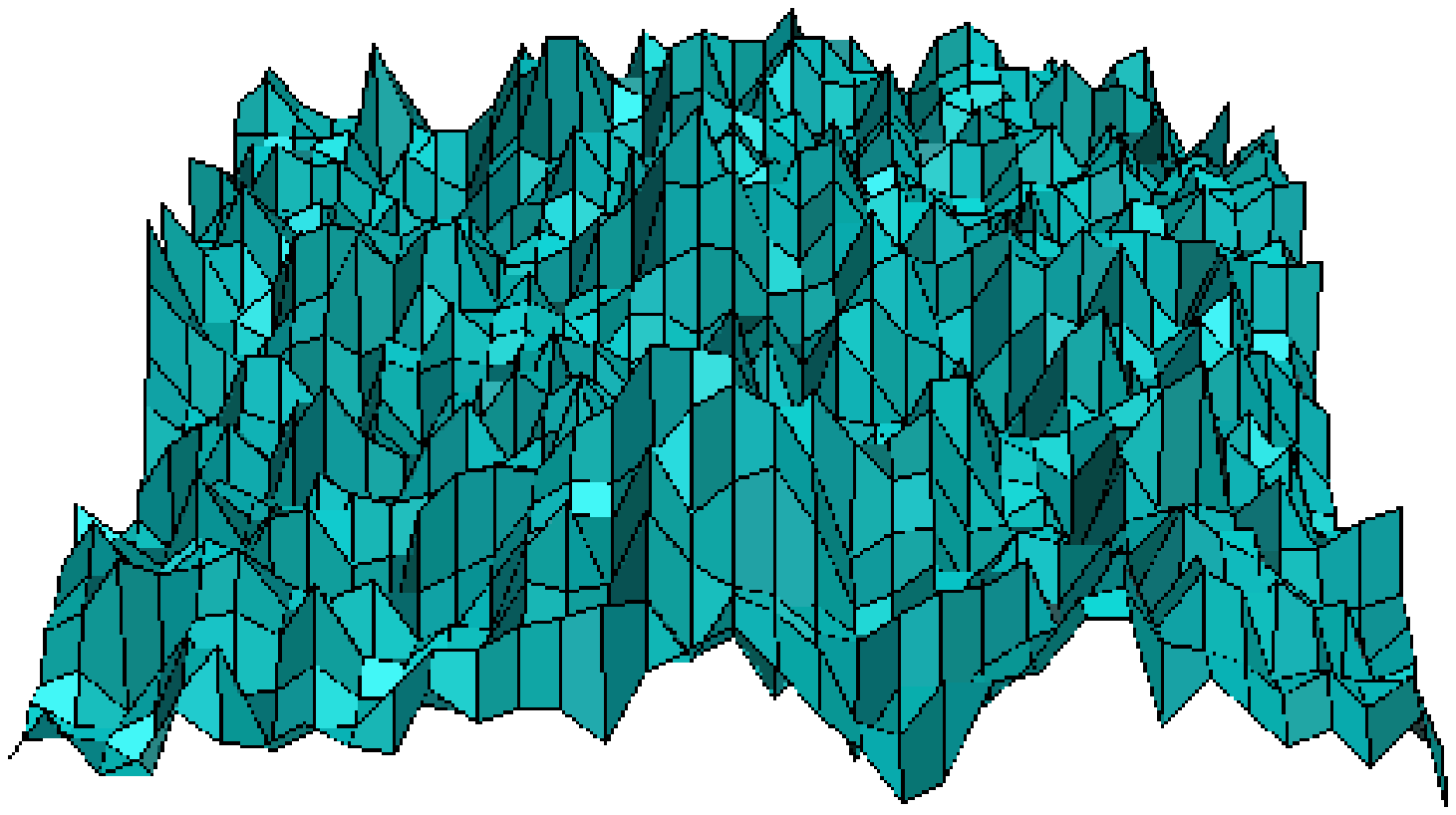}
    \caption{A typical $N=35$ $X_\mu(\sigma)$ ($\mu=1$) configuration
after the gauge fixing.}
\label{fig:f05}
}

\FIGURE{
    \includegraphics[height=4.3cm]{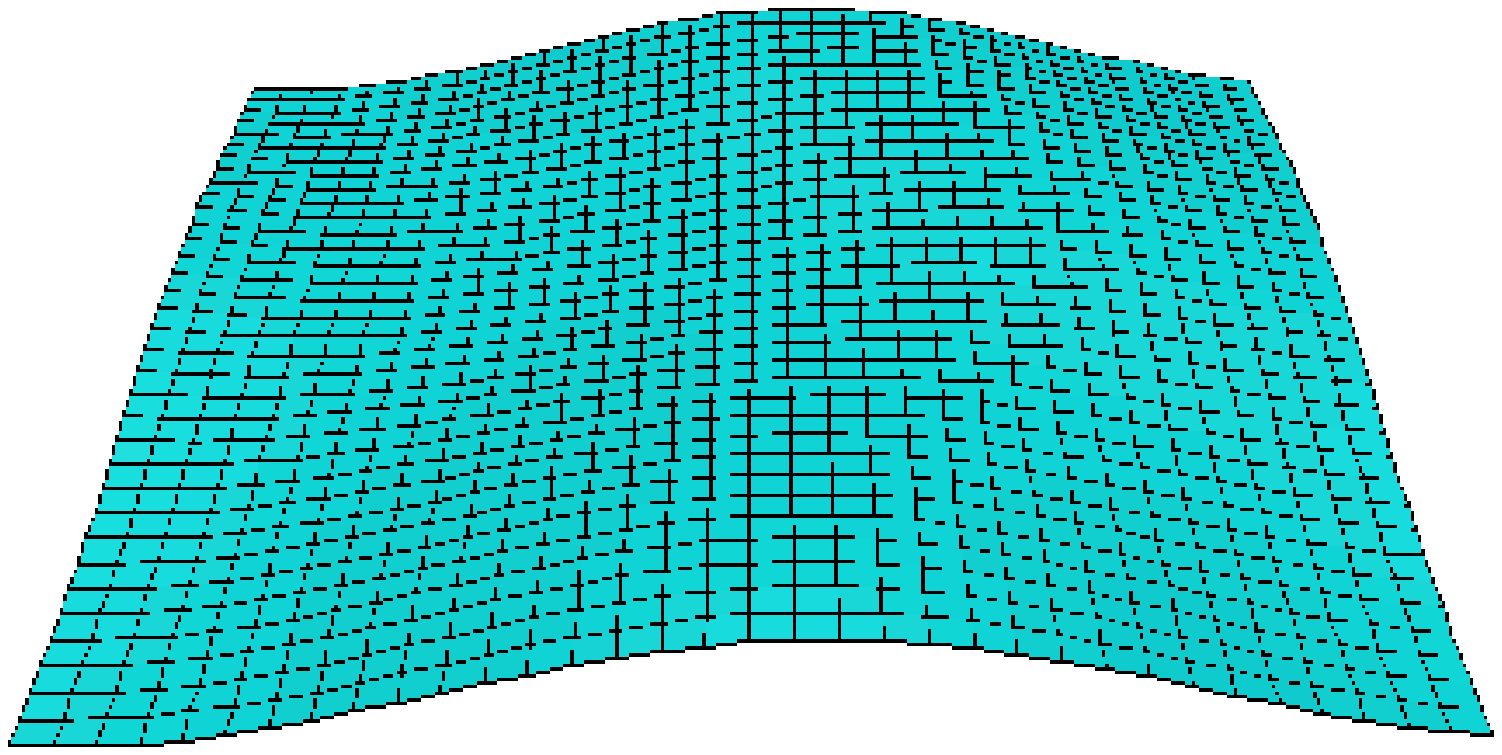}
    \includegraphics[height=4.3cm]{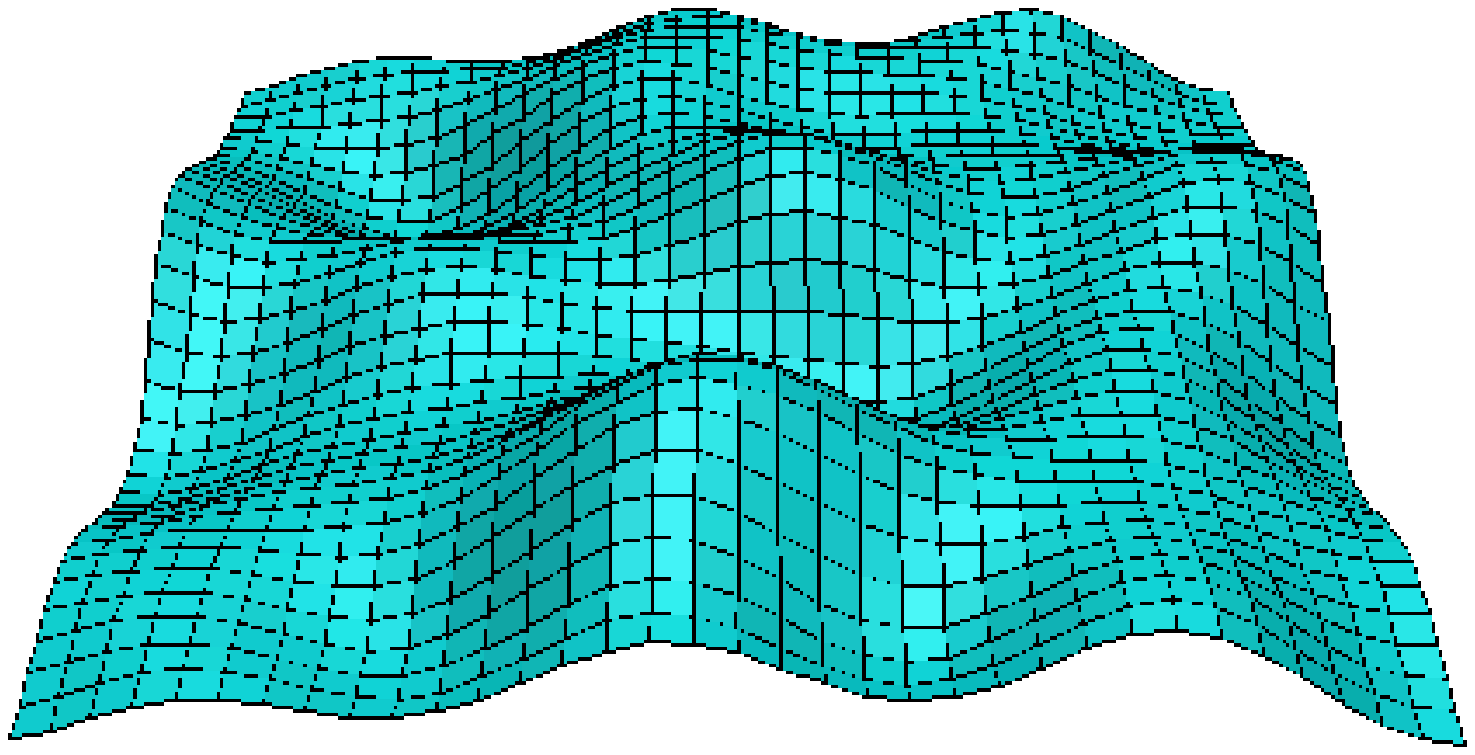}\\
    \includegraphics[height=4.3cm]{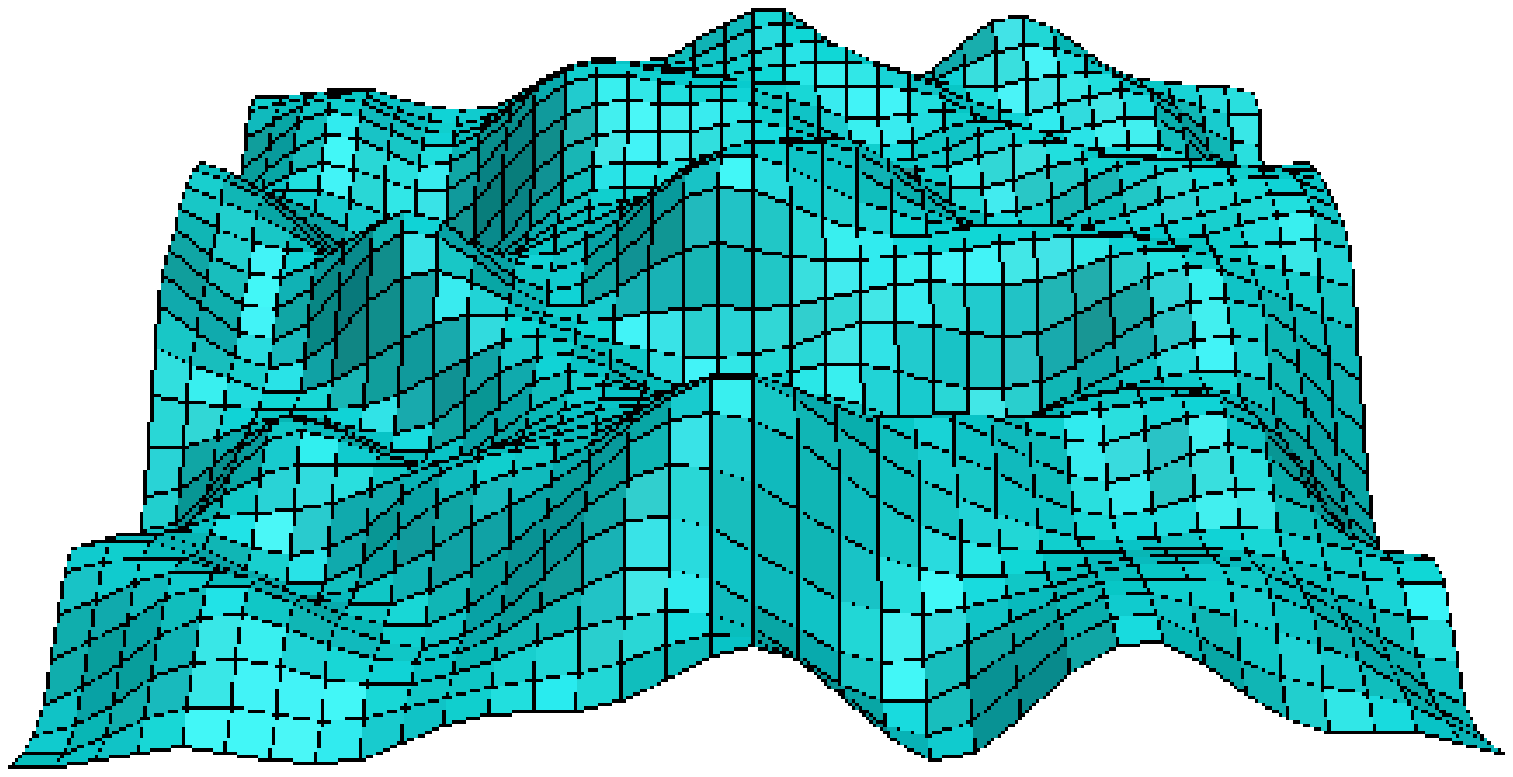}
    \caption{The effect of cutting off Fourier modes higher
    than $k_{\rm c}$ on the
    worldsheet of Fig.~\protect\ref{fig:f05}. On the left $k_{\rm c}=1$,
    on the right $k_{\rm c}=3$ and on the bottom $k_{\rm c}=5$.}
\label{fig:f07}
}

Our numerical calculation starts with generating
configurations of the model (\ref{action})
for $D=4$ and $N=15, 25, 35$ using the method described
in Ref.~\cite{HNT}. 
For each configuration we minimize the
roughness functional $I$ defined by eq.~\rf{roughness}
with respect to the SU($N$) transformation \rf{gauge}.
We perform 2000 sweeps per
configuration, where a sweep is the minimization of $I$
with respect to all SU(2) subgroups of SU($N$)
\cite{HNT,AABHN}. 
{}From the configuration $A_\mu$ 
obtained after the SU($N$) transformation that minimizes
$I$, we calculate through (\ref{Uxinv}) 
the worldsheet configuration $X_\mu (\sigma)$.
When we define an ensemble average $\langle \ \cdot \ \rangle$
in what follows,
we assume that it is taken with respect to $X_\mu (\sigma)$
{\em after} the gauge fixing.
The number of configurations used for an ensemble average
is 658, 100 and 320 for $N=15,25,35$ respectively.
We note that in the present model, finite $N$ effects is known \cite{HNT}
to appear as a $1/N^2$ expansion.
Also, the large $N$ factorization is clearly observed 
for $N=16,32$ \cite{HNT}.
(We have checked that it occurs for $N=15,25,35$ as well.)
We therefore consider that the $N$ we use in the present work
is sufficiently large to discuss the large $N$ asymptotics.

\FIGURE[hbt]{
    \includegraphics[height=9cm]{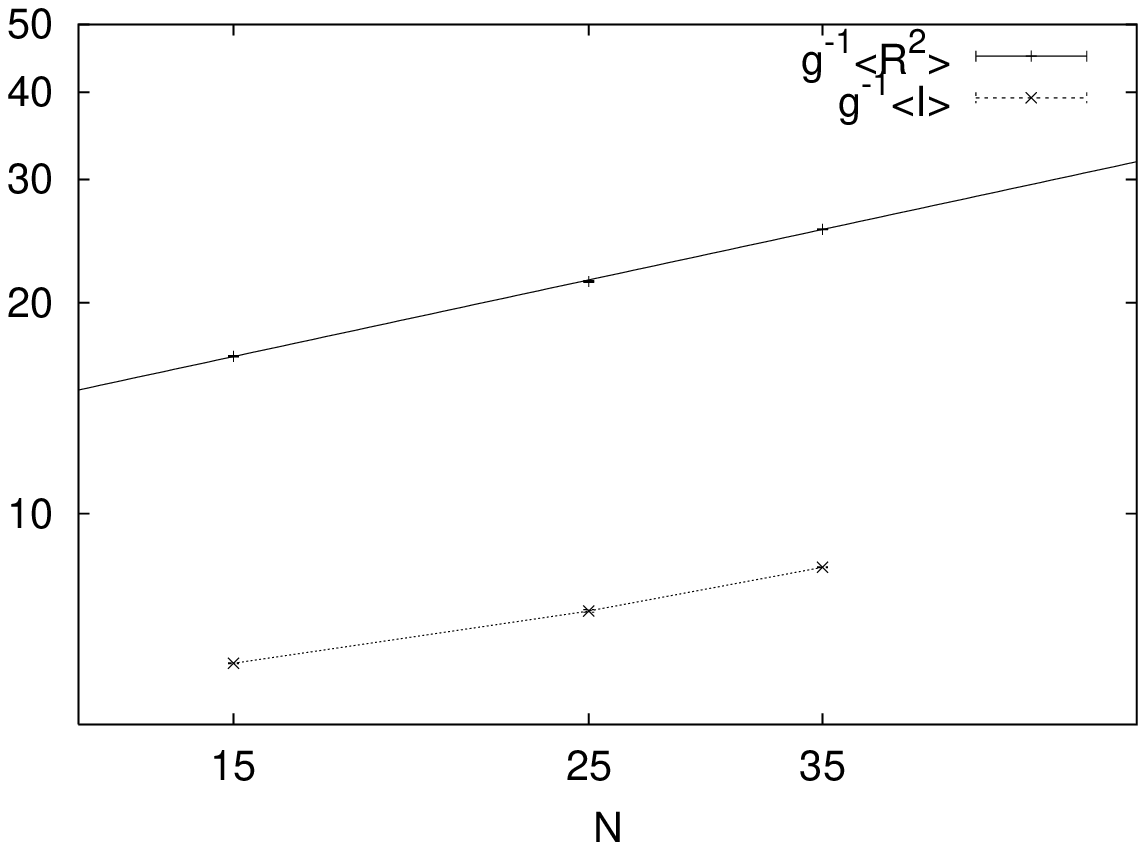}
    \caption{The fluctuation of the surface $R^2$
and the roughness functional $I$, which represents
the average link length in the target space, are plotted against $N$
with the normalization factor $g^{-1}$.
The solid straight line is a fit to the power-law behavior
$\langle R^2 \rangle  \propto g N^{0.493(3)}$.
The dotted line is drawn to guide the eye.
}
\label{fig:Imin-trA}
}

We also note that the parameter $g$, 
which appears in the action (\ref{Moyalaction}),
can be absorbed by the field redefinition 
$X_\mu ' (\sigma)  = \frac{1}{\sqrt{g}} X_\mu (\sigma)$.
Therefore, $g$ is merely a scale parameter,
and one can determine the $g$ dependence of all the observables 
on dimensional grounds.
The results will be stated in such a way that 
they do not depend on the choice of $g$.


\FIGURE[hbt]{
    \includegraphics[height=9cm]{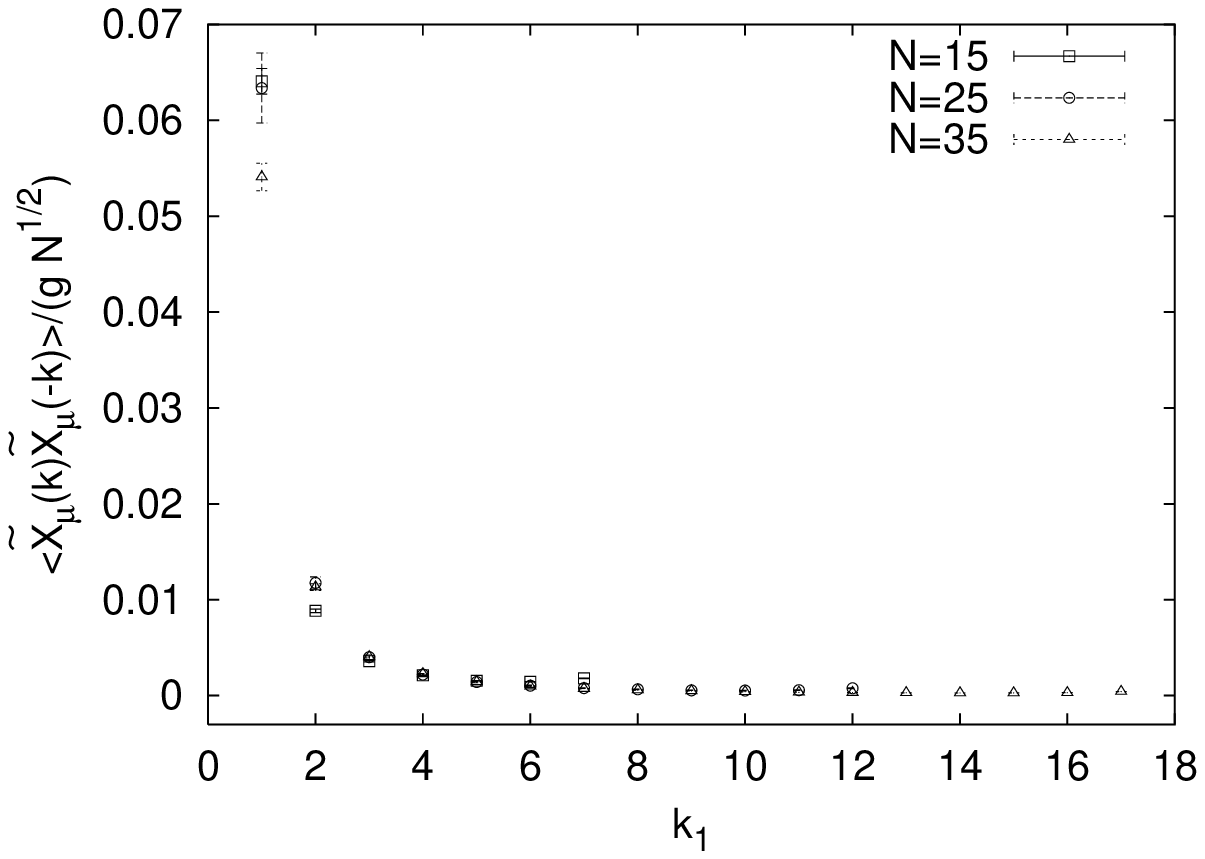}
    \caption{The Fourier mode 
    amplitudes $\vev{\tilde{X}_\mu
    (k)\tilde{X}_\mu (-k)}/(g N^{1/2})$ 
    are plotted against $k_1$ ($k_2 = k_1$)
    for $N=15,25,35$. 
}
\label{fig:f01}
}
\FIGURE[hbt]{
    \includegraphics[height=9cm]{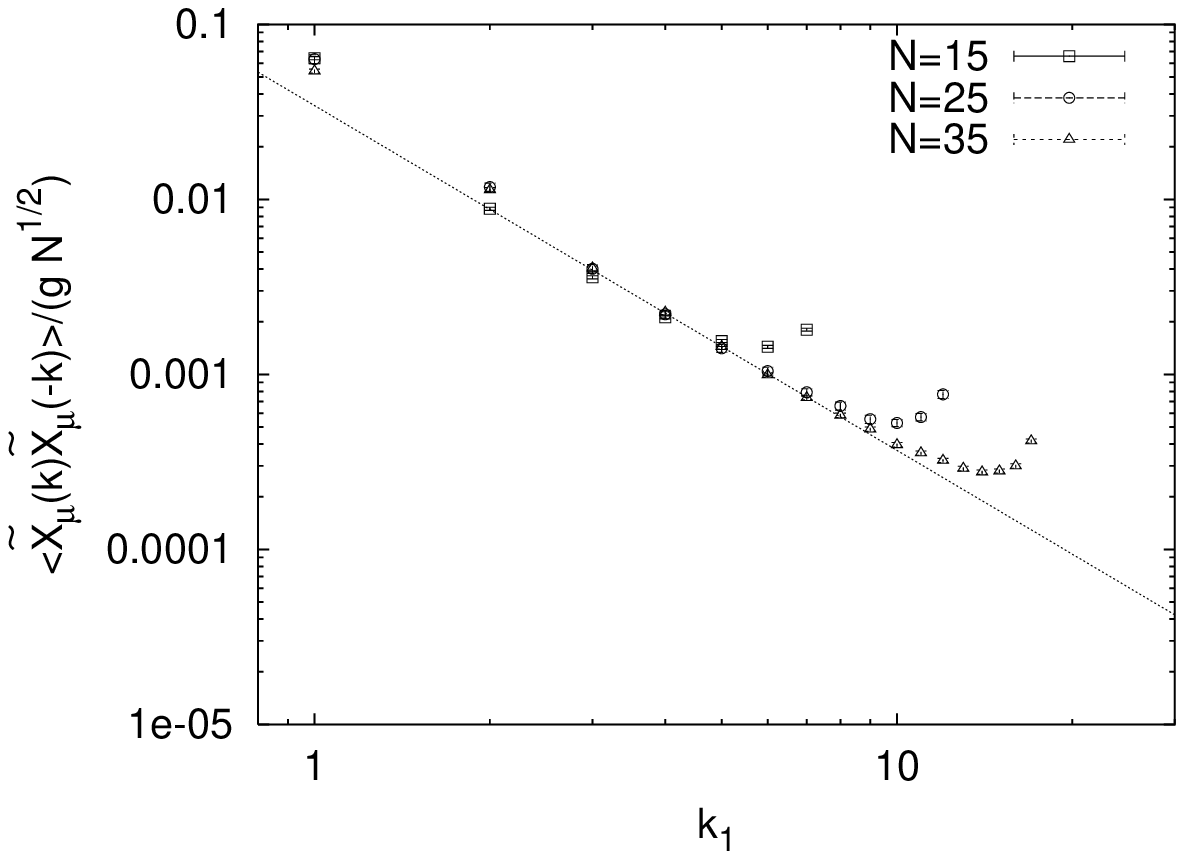}
    \caption{The Fourier mode amplitudes $\vev{\tilde{X}_\mu
    (k)\tilde{X}_\mu (-k)}/(g N^{1/2})$ in Fig.~\ref{fig:f01}
    are now plotted in the log-log scale in order to visualize
    the power-law behavior. 
    The straight line is a fit to $C |k|^{-q}$ with $q=1.96(5)$ 
    for the $N=35$ data.
}
\label{fig:f02}
} 

In Fig.~\ref{fig:f05} we show a typical worldsheet
configuration for $N=35$ after the gauge fixing.
We observe that the worldsheet has no spikes.
We compute the Fourier modes
$\tilde{X}_\mu (k)$ of $X_\mu (\sigma)$ through
\beq
\tilde{X}_\mu (k) 
 \defeq  
\frac{1}{N^2} \sum_{\sigma_i \in \IZs _N} X_\mu (\sigma) 
\eex ^{- 2 \pi i k \cdot \sigma /N} 
=  \frac{1}{N} \tr (A_\mu J_{k}) \ ,
\label{Xtildek}
\eeq
where the range of $k_1$ and $k_2$ are chosen to be $-(N-1)/2$
to $(N-1)/2$.
Fig.~\ref{fig:f07} describes how the same worldsheet configuration
shown in Fig.~\ref{fig:f05}
looks like if we cut off\footnote{More precisely, 
we keep the modes $\tilde{X}_\mu (k)$
with $k_1 \le k_{\rm c}$ and $k_2 \le k_{\rm c}$ and set the other
modes to zero by hand.} the Fourier modes higher than $k_{\rm c}$.
We find that the configuration obtained by keeping 
only a few lower Fourier modes
already captures the characteristic behavior of the original configuration.
We have checked that this is {\em not} the case if we do not fix the
gauge.

We measure the fluctuation of the surface, which is
given by\footnote{We note that the action (\ref{action})
is invariant under $A_\mu \mapsto A_\mu + \alpha_\mu \id_N$.
Therefore, the trace part of $A_\mu$ is completely decoupled
from the dynamics. We fix these degrees of freedom by imposing
the matrices $A_\mu$ to be traceless.
In the language of the worldsheet theory (\ref{Moyalaction}),
the symmetry corresponds to the translational invariance
$X_\mu (\sigma)\mapsto X_\mu (\sigma) + \mbox{const.}$.
The tracelessness condition for $A_\mu$ maps to
$\sum_{\sigma _i \in \IZs _N}  X_\mu (\sigma) = 0$.
}
\beq
R^2 \defeq
\frac{1}{N^4}\sum_{\sigma _i , \sigma ' _i \in \IZs _N}
\Bigl\{  X_\mu (\sigma) -   X_\mu (\sigma ') \Bigr\}^2
= \frac{2}{N^2}\sum_{\sigma _i \in \IZs _N}  X_\mu (\sigma)^2 
= \frac{2}{N}{\rm tr}(A_\mu^2)  \ .
\label{fluctuation}
\eeq
The average of the fluctuation is finite for finite $N$,
and it is plotted in Fig.~\ref{fig:Imin-trA}.
The power-law fit to the large $N$ behavior 
$\langle R^2  \rangle \sim g N^{p}$
yields $p=0.493(3)$,
which is consistent with the result 
$p = 1/2$ obtained in Ref.~\cite{HNT}.
Although the finiteness of the fluctuation already implies a certain
stability of the worldsheet,
we note that the fluctuation defined by the l.h.s.\ of (\ref{fluctuation})
is actually invariant under the star-unitary 
transformation (\ref{stargaugetr}).
In particular, the fluctuation is finite even before the gauge fixing.
Therefore, the smoothness of $X_\mu (\sigma)$
(which appears only after the gauge fixing)
is a notion which is stronger
than the finiteness of the fluctuation.

Let us point out also that the roughness functional $I$ 
actually represents 
the average link length in the target space.
We therefore plot $\langle I \rangle$ in Fig.~\ref{fig:Imin-trA} 
and compare it with the fluctuation (\ref{fluctuation}).
The former is observed to be smaller than 
the latter\footnote{If we do not fix the gauge,
we observe that the two quantities $\langle R^2 \rangle$ 
and $\langle I \rangle$ coincide, meaning that the $X_\mu (\sigma)$
is completely rough.}
which is consistent with the observed smoothness of $X_\mu (\sigma)$.
The ratio $\vev{I}/\vev{R^2}$ is
0.364(1), 0.338(2), 0.3290(5) for $N=15,25,35$, respectively,
which may be fitted to a power-law behavior 
$\vev{I}/\vev{R^2}\sim N^{-0.120(4)}$.
This indicates a tendency that the worldsheet is getting 
smoother as $N$ increases.

\FIGURE[hbt]{
    \includegraphics[height=9cm]{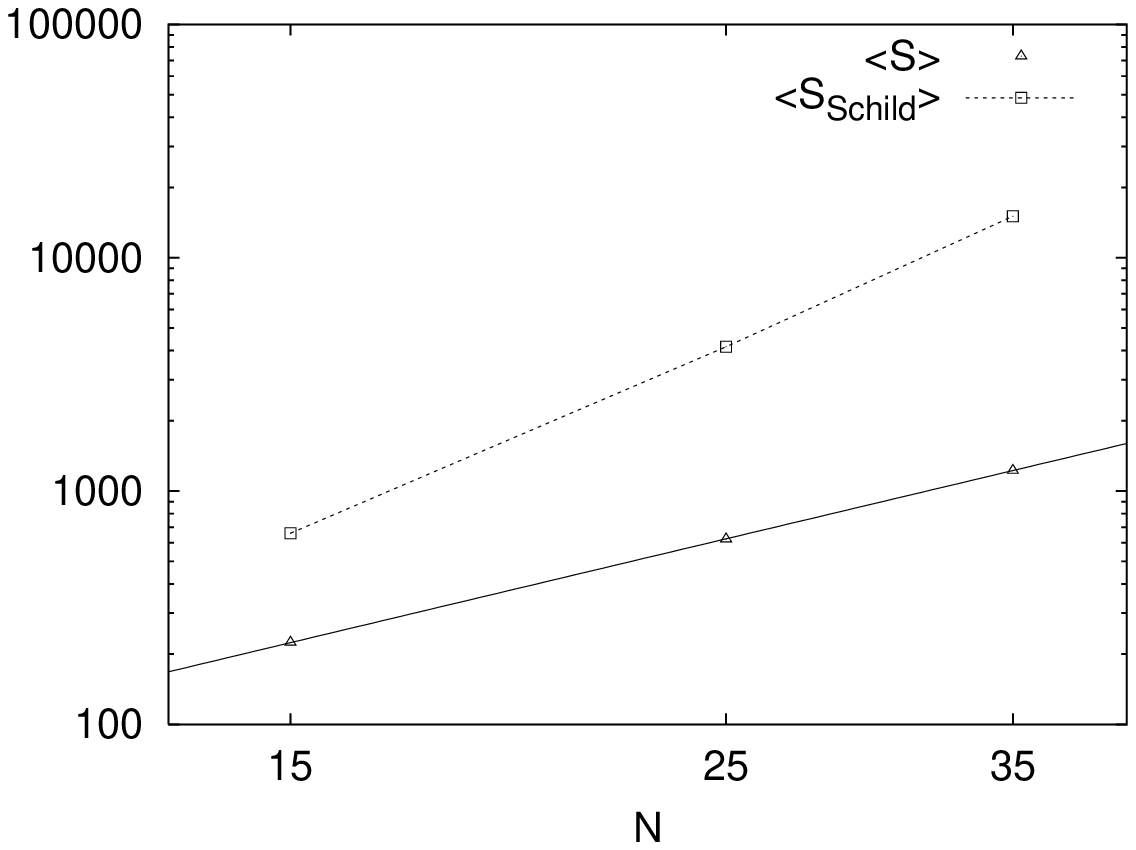}
    \caption{Plot of 
$\vev{S}$ and $\vev{S_{\rm Schild}}$ 
as a function
    of $N$. The solid line represents the exact result
    $\langle S \rangle =(N^2-1)$.
The dotted line is drawn to guide the eye.}
\label{fig:f18}
}

\FIGURE[hbt]{
    \includegraphics[height=9cm]{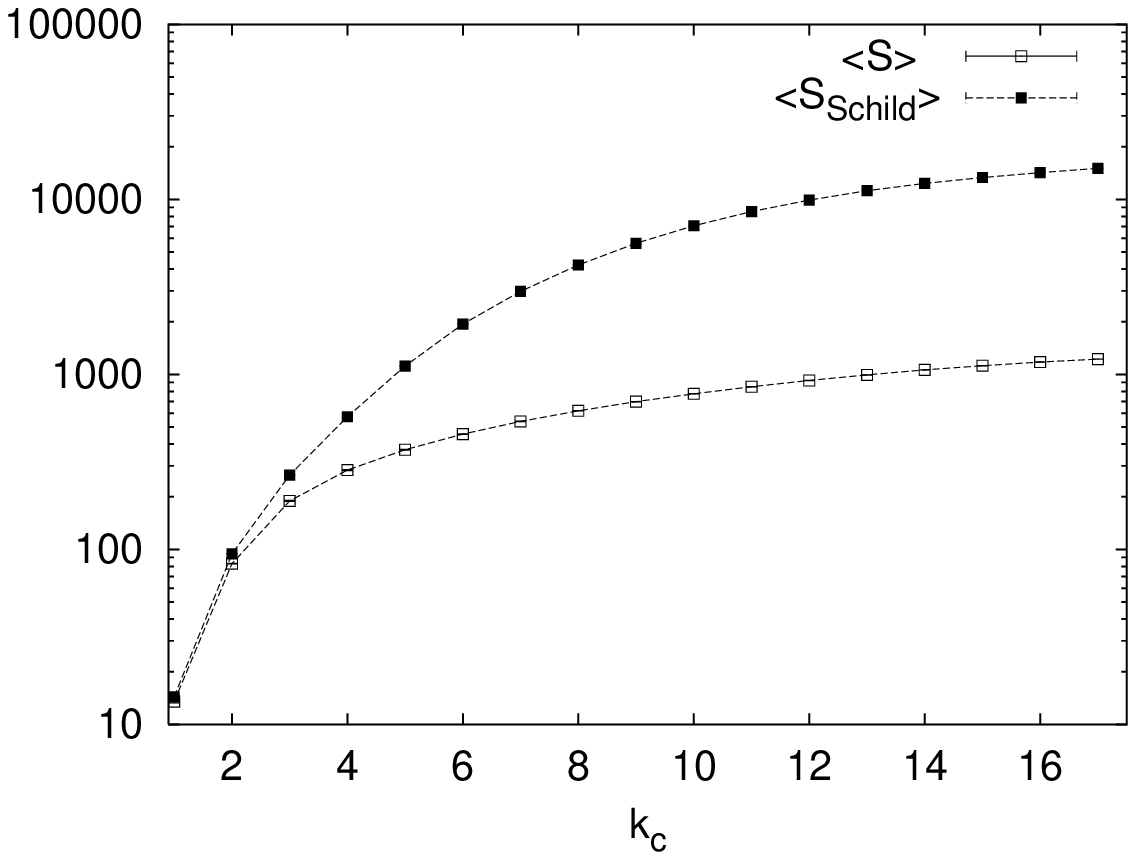}
    \caption{The two actions $\vev{S_{\rm Schild}}$ 
and $\vev{S}$ are computed by cutting off the Fourier modes
higher than $k > k_{\rm c}$ by hand.
The results are plotted as a function of the 
mode cutoff $k_{\rm c}$ for $N=35$.}
\label{fig:f14}
}

In order to quantify the smoothness of $X_\mu(\sigma)$ further,
let us examine the Fourier mode amplitudes.
We first note that there is a relation
\beq
\sum_{k _i \in \IZs _N} \vev{\tilde{X}_\mu(k)\tilde{X}_\mu (-k)} = 
\vev{\frac{1}{N^2}\sum _{\sigma _i \in \IZs _N} 
X_\mu (\sigma)^2} \sim \mbox{O}(g  N^{1/2})  \ .
\label{sumrule}
\eeq
This motivates us to plot 
$\vev{\tilde{X}_\mu(k)\tilde{X}_\mu (-k)}/(g  N^{1/2})$
with this particular normalization.
The results are
shown in Fig.~\ref{fig:f01} and Fig.~\ref{fig:f02}.
We find a good scaling in $N$;
data points for different $N$ fall on top of each other.
The discrepancy in the large $k$ region can be understood
as a finite $N$ effect.
The $k$ dependence of the Fourier mode amplitudes
suggests that the higher modes are indeed suppressed.
Moreover we find that there exist 
a power-law behavior\footnote{If we do not fix the gauge,
we observe that the r.h.s.\ of (\ref{higherKsuppress})
is replaced by a constant independent of $k$, 
which means that the worldsheet behaves like a white noise.
The constant is proportional to $1/N^2$, as expected from 
the relation (\ref{sumrule}), which is gauge independent.
Note, in particular, that the scaling behavior observed
in Fig.~\ref{fig:f02} emerges only after the gauge fixing.}
\beq 
\frac{1}{g N^{1/2}}
\vev{\tilde{X}_\mu(k)\tilde{X}_\mu (-k)}\sim \mbox{const.} \cdot |k|^{-q}
 \ ,
\label{higherKsuppress}
\eeq 
where $|k| = \sqrt{(k_1)^2 + (k_2)^2}$.
Assuming that the constant coefficient on the r.h.s.\ of 
(\ref{higherKsuppress}) is independent of $N$, as suggested
by the observed scaling in $N$,
the power $q$ must be $q>2$
in order that the sum on the l.h.s.\ of (\ref{sumrule})
may be convergent in the large $N$ limit.
The power $q$ extracted from $N=35$ data is $q=1.96(5)$,
which may imply that we have not reached sufficiently large
$k$ (due to the finite $N$ effect mentioned above) 
to extract the correct power.
Although we have seen that the amplitudes of 
the higher Fourier modes are suppressed,
we should remember that their number for fixed $|k|$ 
grows linearly with $|k|$.
Therefore, the higher modes can still be non-negligible.

Let us turn to the question whether the action $S$ 
in (\ref{Moyalaction})
approaches the
Schild action $S_{\rm Schild}$ in 
(\ref{Schildaction})
in the large $N$ limit. 
In terms of Fourier modes, the two actions 
read
\beqa
 \label{k01}
 S 
   &=& - \frac{N}{g^2} \sum_{klp}
       \tilde{X}_\mu(k)\tilde{X}_\nu(l)\tilde{X}_\mu(p)\tilde{X}_\nu(-k-l-p)
       \nonumber\\
 \label{k02}
   & &\quad \times \sin\left(\frac{2\pi}{N}\epsilon_{nm}k_nl_m\right)
                   \sin\left(\frac{2\pi}{N}\epsilon_{rs}p_r(k+l)_s\right)
                \, , \\
 S_{\rm Schild} 
     &=&
     -\frac{N}{g^2 } \left( \frac{N}{2 \pi} \right)^2  \sum_{klp}
      \tilde{X}_\mu(k)\tilde{X}_\nu(l)\tilde{X}_\mu(p)\tilde{X}_\nu(-k-l-p)
    \nonumber\\
     & &\quad\times
      \epsilon_{ij}\sin\left(\frac{2\pi}{N}k_i\right)
                   \sin\left(\frac{2\pi}{N}l_j\right)
      \epsilon_{rs}\sin\left(\frac{2\pi}{N}p_r\right)
                   \sin\left(\frac{2\pi}{N}(k+l+p)_s\right) \ .
 \label{k04}
\eeqa
If the higher Fourier modes can be neglected one can see from
eqs.~\rf{k02} and \rf{k04} that $S = S_{\rm Schild}$ 
in the large $N$ limit. We measure both quantities and plot the results in
Fig.~\ref{fig:f18}. 
The average of the action (\ref{Moyalaction})
is known \cite{HNT} analytically $\langle S \rangle = N^2 -1$,
which is clearly reproduced from our data.
On the other hand, $\langle S_{\rm Schild}\rangle$ is much larger,
and moreover it grows much faster, the growth being close to O($N^4$).
Therefore we can safely conclude that the two quantities 
do not coincide in the large $N$ limit.

The disagreement of the two actions
(\ref{Moyalaction}) and (\ref{Schild}) in the large $N$ limit
implies that the higher modes play a crucial role.
In order to see their effects explicitly,
we cut off the Fourier modes higher than $k_{\rm c}$.
In Fig.~\ref{fig:f14},
we plot the average of the actions thus calculated
against $k_{\rm c}$ for $N=35$.
The two actions with the cutoff at $k_{\rm c}$
are indeed identical for small $k_{\rm c}$,
but they start to deviate from each other as $k_{\rm c}$ increases,
ending up with totally different values at $k_{\rm c} = (N-1)/2$,
i.e. when all the modes are included.

Finally, let us discuss the extrinsic geometry of the worldsheet.
One may define the Hausdorff dimension $d_H$ of the worldsheet through
\beq
\frac{\langle A\rangle}{\ell ^2}  
\propto \left( \frac{\sqrt{\langle R ^2\rangle}}{\ell}  \right)^{d_H}
\mbox{~~~~~~as~}N\rightarrow \infty \ ,
\label{Hausdorff}
\eeq
where $R$, which is defined by eq.~\rf{fluctuation},
represents the extent of the worldsheet in the target space.
We have defined $A$, the total area of 
the worldsheet in the target space, by
\beq
 A =  a ^2
 \sum_{\sigma_i \in \IZs _N}
 \sqrt{  \{   X_\mu(\sigma), X_\nu(\sigma) \}   ^2 }  \ ,
\label{totalarea}
\eeq
where the Poisson bracket $\{ \, \cdot \, \}$ is defined by
(\ref{latticePoisson}) and
$a = \frac{2 \pi}{ N} $ is the lattice spacing
on the worldsheet. Eq.~(\ref{totalarea}) is nothing but the 
Nambu-Goto action.
The scale parameter $\ell$ in eq.~\rf{Hausdorff}
should be introduced in order to make the equation 
consistent dimensionally.
A natural choice of the fundamental scale $\ell$ is
the average link length in the target space,
which is represented by $\sqrt{\langle I\rangle}$,
where $I$ is the roughness functional defined by eq.~\rf{roughness}.
We observe, as expected, that $\langle A\rangle \sim N^2 \langle I\rangle$,
which means that the l.h.s.\ of (\ref{Hausdorff}) is of O($N^2$).
As we have seen in Fig.~\ref{fig:Imin-trA}, 
we observe that $\vev{I}/\vev{R^2}\sim N^{-0.120(4)}$.
This means that the Hausdorff dimension $d_H$ defined by
(\ref{Hausdorff}) with the choice $\ell = \sqrt{\langle I\rangle}$
is $d_H \sim 33$, which might suggest that actually $d_H = \infty $.
Therefore, the extrinsic geometry of the embedded worldsheet is 
described by a crumpled surface.

\section{Conclusions}
\label{conclusions}

In this paper we have studied the star-Schild action \rf{nonc}
resulting from the zero-volume limit of SU($N$) Yang-Mills theory. 
We find that the star-Schild action does not approach
the Schild action \rf{sch} due to the important role played by
the ever-increasing number of higher modes. This has some
implications to the
ideas presented long ago \cite{longtime} that it might be that the
Schild action represents the large $N$ QCD action. From our results
the two actions differ more and more with increasing $N$.
The Poisson bracket increases much faster with
$N$ than the corresponding Moyal bracket. 
Our conclusion is therefore that QCD strings would be described
by a noncommutative string theory 
defined by the star-Schild action,
rather than the standard Schild action.

As we have seen, it is possible to find a star-unitary transformation 
$g(\sigma)$ such that the 
surfaces defined by the star-Schild action are
regular (i.e., do not have long spikes), and this action therefore 
defines a new type of string theory. The theory 
is invariant under star-unitary transformations, which generalize the
area-preserving diffeomorphisms, the invariance of 
the usual Schild action.
As we discuss in the Appendix \ref{append_diffeo},
the reparametrization invariance of the
worldsheet fields $X_\mu (\sigma)$ is
restricted to linear transformations of the $\sigma$'s
in the case of the star-Schild action.
(Note, however, that the reparametrization invariance 
of the usual Schild action is also much reduced relative
to the Nambu-Goto action.)
The star-unitary transformations transform the worldsheet 
configuration in such a way that the changes cannot be absorbed by
a reparametrization.
In obtaining a regular surface 
we have chosen a particular ``gauge''. 
The regularity will not be changed drastically
under smooth star-unitary transformations.
However, if we do not fix the gauge, we obtain
spiky surfaces, which is connected to a regular surface
by a rough star-unitary transformation $g(\sigma)$. 
The main point is that it is at all
{\it possible} to obtain a regular surface by fixing the gauge properly.

A possible intuitive understanding of the regularity of the worldsheet
in the noncommutative string is 
that the action contains higher derivatives in
the star-product. Let us recall that one of the
motivations for introducing extrinsic curvature (which also contains
higher derivatives in a different combination) was \cite{ec} that this
extra term in the action makes the worldsheet more stiff. 
One would also expect a similar effect 
from the introduction of higher
derivatives, because an extremely rough worldsheet with long spikes
would have at least
some derivatives rather large. Although the star-product contains these
derivatives in a special combination, it is difficult for all these large 
derivatives to cancel, and hence a surface dominated by long spikes
would not be preferred.
This is confirmed by the observation that the average link length
is much smaller than the average extent.
The extrinsic geometry of the embedded surface, on the other hand, 
is described by a crumpled surface with a large Hausdorff dimension.

It would be of interest to address the issues studied in this paper
in the supersymmetric case
using the numerical method developed in Ref.~\cite{AABHN}.
We hope to report on it in a future publication.

\acknowledgments
We thank Jan Ambj\o rn for interesting discussions on instabilities
in bosonic strings. We
are also grateful to Z.~Burda for helpful comments.
K.N.A. acknowledges interesting discussions with E.\ Floratos,
E.\ Kiritsis and G.\ Savvidy. K.N.A.'s research was partially supported
by the RTN grants HPRN-CT-2000-00122 and HPRN-CT-2000-00131 and a
National Fellowship Foundation of Greece (IKY) postdoctoral
fellowship.

\bigskip

\appendix
\section{Derivation of the roughness functional}
\label{append_rough}

In this appendix, we rewrite the roughness functional
(\ref{defI}) in terms of the matrices $A_\mu$ and
derive (\ref{conv}).
For that purpose, we introduce $N \times N$ unitary matrices
\beqa
D _1 &=& (\Gamma _2 ^\dag)^{\frac{N-1}{2}}  \\
D _2 &=& (\Gamma _1)^{\frac{N-1}{2}} \ ,
\eeqa
which satisfy
\beq
D_j \Gamma _i D_j ^\dag = \eex ^{- 2 \pi i \delta_{ij}/N} \Gamma _i  \ .
\eeq
One can check that
\beq
D_j \Delta (\sigma ) D_j  ^\dag = \Delta (\sigma - \hat{j}) \ ,
\eeq
which implies
\beq
X_\mu (\sigma + \hat{j}) =
\frac1N\,\tr\Bigl(D_j A_\mu D_j  ^\dag 
\Delta(\sigma )\Bigr) \ .
\eeq
Thus the matrix $D_j$ plays the role of a shift operator in the
$j$-direction.
Now we can rewrite the roughness functional $I$ in terms of the
matrices $A_\mu$ as
\beqa
I  &=&  \frac{1}{2 N}  \sum_{j} 
\tr (  D_j A_\mu D_j  ^\dag  - A_\mu )^2 \n
&=& \frac{1}{2 N}  \sum _{IJ} \sum_{j}
\left| ( D_j A_\mu D_j  ^\dag)_{IJ}  - (A_\mu )_{IJ} \right|^2  \ .
\eeqa
Using the explicit form of $\Gamma _\mu$
given by eq.~(\ref{clockshift}), we obtain
\beqa
(D_1 A_\mu D_1  ^\dag)_{IJ} &=& 
(A_{\mu})_{IJ} \eex ^{2\pi i(I-J)/N} \\
(D_2 A_\mu D_2  ^\dag)_{IJ} &=& 
(A_{\mu})_{I+\frac{N-1}{2},J+\frac{N-1}{2}} \ ,
\eeqa
which yields (\ref{conv}).

\section{Star-unitary invariance and area-preserving diffeomorphisms}
\label{append_diffeo}

In this appendix, we discuss the relationship between
the symmetry of the Schild action and that of the star-Schild action.
Here only we consider that the worldsheet is given by an 
infinite two-dimensional flat space parametrized
by the continuous variables $\sigma _1$ and $\sigma _2$.
Let us define the Schild and star-Schild actions
\beqa
I_1 &=& \int \ddd ^2 \sigma \, \{  \phi_1 (\sigma) , \phi_2 (\sigma) \} ^2 \\
I_2 &=& \int \ddd ^2 \sigma \,  
\{ \!\! \{ \phi_1 (\sigma) ,  \phi_2 (\sigma) \} \!\! \} _\theta  ^2 \ ,
\eeqa
where the Poisson and Moyal brackets are defined by
\beqa
\{  \phi_1 (\sigma) , \phi_2 (\sigma) \} &=& \epsilon _{ij} 
\frac{\del \phi_1}{\del \sigma _i} \frac{\del \phi_2}{\del \sigma _j} \\
\{ \!\! \{ \phi_1 (\sigma) ,  \phi_2 (\sigma) \} \!\! \}_\theta   &=& 
\frac{1}{\theta} \phi_1 (\sigma) \sin (\theta \epsilon _{ij}  
\overleftarrow{\del_i} 
\overrightarrow{\del_j} ) \phi_2 (\sigma) \ .
\eeqa

The Schild action $I_1$ is invariant under the area-preserving
diffeomorphism
\beq
\sigma_i \mapsto \sigma_i + \epsilon _{ij} \del _j f(\sigma) \ ,
\label{coordinatetr}
\eeq
where $f(\sigma)$ is some infinitesimal real function of $\sigma$.
Under the infinitesimal area-preserving diffeomorphism,
the fields transform as a scalar $\phi_\alpha (\sigma) 
= \phi '_\alpha (\sigma')$,
so that one can state the invariance as the one under
the field transformation
\beq
\phi_\alpha (\sigma) \mapsto 
\phi_\alpha (\sigma) + \{ f(\sigma) , \phi_\alpha (\sigma) \} \ .
\label{PBtransform}
\eeq
On the other hand, the star-Schild action $I_2$ is invariant under
the star-unitary transformation
\beq
\phi_\alpha (\sigma) \mapsto \phi_\alpha (\sigma) 
+ \{ \!\! \{ f (\sigma) ,  \phi_\alpha (\sigma) \} \!\! \}_\theta  \ .
\label{MBtransform}
\eeq
Obviously, this transformation (\ref{MBtransform})
reduces to 
(\ref{PBtransform}) if $\phi _\alpha (\sigma)$ and 
$f(\sigma)$ do not contain
higher Fourier modes compared with $\theta ^{-1/2}$.

In general the two transformations
(\ref{PBtransform}) and (\ref{MBtransform}) differ.
However, we note that they become identical if $f(\sigma)$
contains terms only up to quadratic in $\sigma$ as
\beq
f(\sigma) = a_i \sigma_i + b _{ij} \sigma_i \sigma_j \ ,
\eeq
where $b_{ij}$ is a real symmetric tensor.
{}From (\ref{coordinatetr}), one finds that the 
corresponding coordinate transformation is
\beq
\sigma_i \mapsto \sigma_i + (v_i + \lambda _{ij}\sigma_j ) \ ,
\eeq
where $v_i = \epsilon _{il} a _l$ and
$\lambda_{ij} = \epsilon _{il} b_{lj}$,
which is traceless.
This transformation
includes the Euclidean group, namely translation and rotation. 
Thus we find that the linear (finite) transformation of the coordinates
$\sigma ' _ i = \Lambda _{ij} \sigma_j + v_i$, 
where $\det \Lambda = 1$,
can be expressed as a star-unitary transformation.
In other words, the reparametrization invariance of the 
star-Schild action is restricted to
such linear transformations.




\listoftables           
\listoffigures          


\begin{thebibliography}{99}
\bibitem{alvarez} O.\ Alvarez, \prd{24}{1981}{440};
M.E.\ Cates, {\em Europhys.\ Lett.\ } {\bf 8} (1988) 656.
\bibitem{dt} J.\ Ambj\o rn, B.\ Durhuus, and J.\ Fr\"ohlich, 
\npb{257}{1985}{433}; J.\ Ambj\o rn, B.\ Durhuus, J.\ Fr\"ohlich, and
P.\ Orland, \npb{270}{1986}{457}.\\
For a recent review, see
J.\ Ambj\o rn, B.\ Durhuus and T.\ Jonsson,
``Quantum Geometry'', Cambridge Monographs, 1997.
\bibitem{poles}
S.\ Oda and T.\ Yukawa, \ptp{102}{1999}{215} 
[\hepth{9903216}]; 
P.\ Bialas, Z.\ Burda, B.\ Petersson and J.\ Tabaczek, 
\npb{592}{2000}{391},
[\heplat{0007013}].
\bibitem{ec} 
W.\ Helfrich, \jp{46}{1985}{1263};
L.\ Peliti and
S.\ Leibler, \prl{54}{1985}{1690};
D.\ F\"orster, \pla{114}{1986}{115};
A.\ Polyakov, \npb{268}{1986}{406}.
\bibitem{exx} J.\ Ambj\o rn, A.\ Irb\"ack, J.\ Jurkiewicz, and B.\
Petersson, \npb{393}{1993}{571}, [\heplat{9207008}];
K.N.\ Anagnostopoulos, M.\ Bowick, P.\ Coddington, M.\ Falcioni, L.\ Han,
G.\ Harris, E.\ Marinari, \plb{317}{1993}{102}, [\hepth{9308091}];
J.\ Ambj\o rn, Z.\ Burda, J.\ Jurkiewicz, and B.\ Petersson,
\plb{341}{1995}{286}, [\hepth{9408118}];  
\bibitem{KS} 
W.\ Krauth and M.\ Staudacher,
\plb{435}{1998}{350},
[\hepth{9804199}].
\bibitem{HNT} T.~Hotta, J.~Nishimura and A.~Tsuchiya, 
\npb{545}{1999}{543}, [\hepth{9811220}].
\bibitem{Austing:2001bd}
P.~Austing and J.~F.~Wheater,
hep-th/0101071.




\bibitem{longtime} J.\ Hoppe, \ijmpa{4}{1989}{5235};
E.G.\ Floratos and J.\ Iliopoulos, \plb{201}{1988}{237}; 
E.G.\ Floratos, J.\ Iliopoulos, and G.\ Tiktopoulos,
\pl{217}{1989}{217}; 
D.\ Fairlie, P.\ Fletcher, and C.\ Zachos, \plb{218}{1989}{203};
D.\ Fairlie and C.\ Zachos, \plb{224}{1989}{101};
D.\ Fairlie, P.\ Fletcher, and C.\ Zachos, \jmp{31}{1990}{1088}; 
I.\ Bars, \plb{245}{1990}{35}; 
E.G.\ Floratos and G.K.\ Leontaris, \plb{464}{1999}{30}, [\hepth{9908106}].
\bibitem{BFSS} T.\ Banks, W.\ Fischler, S.H.\ Shenker and L.\ Susskind,
\prd{55}{1997}{5112},
[{\tt hep-th/9610043}].
\bibitem{IKKT} N.\ Ishibashi, H.\ Kawai, Y.\ Kitazawa and A.\ Tsuchiya, 
\npb{498}{1997}{467}, [\hepth{9612115}]
\bibitem{EK} T.\ Eguchi and H.\ Kawai, \prl{48}{1982}{1063}.
\bibitem{GrossKitazawa}
D.\ Gross and Y.\ Kitazawa, Nucl.\ Phys.\ {\bf B206} (1982) 440.
\bibitem{contEK} A.~Gonz\'alez-Arroyo and C.P.~Korthals Altes,
\plb{131B}{1983}{396}.
\bibitem{quench}
G.\ Bhanot, U.\ Heller and H.\ Neuberger, Phys.\ Lett.\ {\bf 113B} (1982)  
47.
\bibitem{GO} A.~Gonz\'alez-Arroyo and M.\ Okawa, 
Phys.\ Rev.\ {\bf D27} (1983) 2397.
\bibitem{AABHN}J.\ Ambj\o rn, K.N.\ Anagnostopoulos, W.\ Bietenholz,
T.\ Hotta and J.\ Nishimura, \jhep{0007}{2000}{013}, [\hepth{0003208}].
\bibitem{tHooft} G.~'t~Hooft,
``Monopoles, Instantons and Confinement'',
notes written by F.~Bruckmann, [\hepth{0010225}]. 
\bibitem{BM} I.\ Bars and D.\ Minic, \hepth{9910091}.
\bibitem{AMNS}
J.\ Ambj\o rn, Y.M.\ Makeenko, J.\ Nishimura and R.J.\ Szabo,
\jhep{9911}{1999}{029}, [\hepth{9911041}];
\plb{480}{2000}{399}, [\hepth{0002158}];
\jhep{0005}{2000}{023}, [\hepth{0004147}].



\bibitem{matrixmodels}
J.\ Nishimura, 
\mpla{11}{1996}{3049},
[\heplat{9608119}]; 
T.\ Suyama and A.\ Tsuchiya,
\ptp{99}{1998}{321},
[\hepth{9711073}];
%
T.\ Nakajima and J.\ Nishimura,
\npb{528}{1998}{355},
[\hepth{9802082}];
%
W.\ Krauth, H.\ Nicolai and M.\ Staudacher,
\plb{431}{1998}{31},
[\hepth{9803117}];
%
%
%
\plb{453}{1999}{253},
[\hepth{9902113}];
%
%
J.\ Ambj\o rn, K.N.\ Anagnostopoulos, W.\ Bietenholz,
T.\ Hotta and J.\ Nishimura,
\jhep{0007}{2000}{011},
[\hepth{0005147}];
%
%
J.\ Nishimura and G.\ Vernizzi,
\jhep{0004}{2000}{015},
[\hepth{0003223}];
%
\prl{85}{2000}{4664},
[\hepth{0007022}];
%
%
S.\ Horata and H.S.\ Egawa, 
\hepth{0005157};
%
S.\ Oda and F.\ Sugino, 
\hepth{0011175};
%
Z.\ Burda, B.\ Petersson and J.\ Tabaczek, 
\heplat{0012001}.



\bibitem{Kitsune}N.\ Kitsunezaki and S.\ Uehara, \hepth{0010038}.

\end{thebibliography}
\end{document}